\documentclass[%
groupedaddress,
preprint,
 amsmath,amssymb,
 aps,
prx,
]{revtex4-2}

\usepackage{CJK}
\usepackage{times}
\usepackage{hyperref}
\usepackage{color}
\usepackage{amsmath}  
\usepackage{amsfonts} 
\usepackage{graphicx} 
\usepackage{caption}
\usepackage{subfigure}
\usepackage{float}
\usepackage{bm}
\usepackage{amssymb}
\usepackage{marvosym}

\begin{document}
\begin{CJK*}{UTF8}{gbsn}
\raggedbottom
\title{Turing Pattern and Chemical Medium-Range Order of Metallic glasses}

\author{Song Ling Liu (刘松灵)\textsuperscript{1,2}}
\author{Xin Yu Luo (罗新宇)\textsuperscript{1,2}}
\author{Jing Shan Cao (曹景山)\textsuperscript{1,2}}
\author{Zhao Yuan Liu (刘召远)\textsuperscript{3}}
\author{Bei Bei Xu (许贝贝)\textsuperscript{4}}
\author{Yong Hao Sun (孙永昊)\textsuperscript{1,2,5,\Letter}}
\author{Weihua Wang (汪卫华)\textsuperscript{1,2,5}}


\affiliation{1.Institute of Physics, Chinese Academy of Sciences, Beijing, 100190, China.}
\affiliation{2.School of Physical Sciences, University of Chinese Academy of Sciences, Beijing, 100190, China.}
\affiliation{3.Shandong Computer Science Center (National Supercomputer Center in Jinan), Qilu University of Technology (Shandong Academy of Sciences), Jinan, Shandong, 250014, China.}
\affiliation{4.Shanghai Institute of Microsystem and Information Technology, Chinese Academy of Sciences, Shanghai, 200050, China.}
\affiliation{5.Songshan Lake Materials Lab, 523808, Dongguan, Guangdong, China}
\affiliation{\rm \Letter\ Corresponding authors: ysun58@iphy.ac.cn }
\date{\today}
\begin{abstract}
The formation of bulk metallic glass requires the constituent elements to have a negative heat of mixing but has no restrictions on its magnitude. An understanding of this issue is lacking due to the absence of a valid method for describing chemical ordering of metallic glasses. For example, the radial distribution function is ineffective in identifying the elemental preferences of packed atoms. Here, we show that using molecular-dynamics simulation, the chemical medium-range ordering of liquid alloys can be evaluated from persistent homology. This inherently arising chemical medium-range order in metallic glasses is exclusively regulated by the activation and inhibition of the constituent components, making the topology of metallic glasses a Turing pattern. The connecting schemes of atoms of the same species form three distinct regions, reflecting different correlations at the short and medium length scales, while the difference in the schemes corresponds to chemical ordering. By changing the elemental types, it is demonstrated that the chemical medium-range order strongly depends on the relative depth of the interatomic-potential wells. The study separates metallic glasses from crystals under the condition of negative heat of mixing by emphasizing their fundamental difference in interatomic potentials.

\end{abstract}
\maketitle
\end{CJK*}
\clearpage

\section*{Introduction}

According to the empirical rule of Inoue, one of the prerequisites for casting bulk metallic glasses (MGs) is the negative heat of mixing ($\Delta H_{\rm mix}$) between any two elements of the constituents \cite{Tak05}. A negative $\Delta H_{\rm mix}$ suggests a bonding choice preferential to unlike species against like species, so factors like phase separation that are detrimental to glass formation during casting can be avoided. One might infer from this that giant MGs can be produced by lowering $\Delta H_{\rm mix}$ continuously or by choosing elements with $\Delta H_{\rm mix}$ of higher absolute values. Unfortunately, the truth is different: a study of 7688 alloys revealed that the critical cooling rate ($R_{\rm c}$) of binary MGs is only weakly correlated with the absolute value of negative $\Delta H_{\rm mix}$ \cite{Hu2019,Hu2023}.
It is unclear why a negative $\Delta H_{\rm mix}$ is significant but its magnitude has no bearing on MG formation.

Since mixing involves chemical affinity of the elements, the task requires an understanding of the chemical ordering inside the MGs \cite{Ni-Zr2008,Ni-Nb2014}, which implies bonding preference and how far the local chemistry departs from a random melt. Our knowledge, however, is rather limited and largely focused on the first radial-distribution-function (RDF) peak. Although many ordering characteristics describing the structural and chemical order of the first nearest neighbors are proposed \cite{W19,Miracle2004,W21,W215,G14,W38,X-ray1950,X-ray1990,Car1981}, it is unclear what ordering characteristics contribute to the RDF peaks after the first one. Additionally, although the chemical affinity of the solute and solvent atoms of alloys are resembled using a variety of radii by Miracle \cite{Miracle2003,Miracle2004,Miracle2006,Miracle2015} following up with Bernal's hard sphere model \cite{Bernal1959, Bernal1960,NCref2,NCref3,NCref4}, these approaches are still insufficient to resolve the mystery of negative $\Delta H_{\rm mix}$ in the formation of MGs. A model emphasizing the chemical ordering is required, because configurations would change if the affinity of the alloying elements changed even though the radii are set. 
To understand the medium-range ordering (MRO) of amorphous materials, numerous simulation studies have been carried out \cite{H4,H6,H7}, but they all have some drawbacks. The bond-angle or dihedral-angle distribution only highlights configurations of the third-nearest neighbors \cite{dihedral2003,bond2019}. Ring statistics are only useful for polymeric glasses and crystals \cite{H2,H8,H9}. The MRO topology was defined in earlier investigations of persistent homology analysis, but the chemical influence was missed \cite{2016PH,2020SA}. On the other hand, an analogy from DNA helps us to understand the significance of finding CMRO in liquids. While the statistical findings from X-ray diffraction assisted people in understanding the double helix structure of DNA reflecting MRO, the sequence of complementary base-pairing of DNA representing chemical order, lays the foundation for the variation of life. Thus, developing a new model that visualize the chemical medium-range order (CMRO) is what waits to be resolved. 

The current work provides a simple mathematical model that, while ignoring the specifics, draws out the logical connections in complex glass-forming liquids. The concept is derived from Turing's reaction-diffusion theory \cite{turing,science2010}, which explains how ordered patterns spontaneously develop from random complex systems with only an activator and an inhibitor and has been applied in many other fields, such as biology \cite{PNAS2009,PNAS4,PNAS5}, chemistry \cite{np8,np9,np10,np11}, and 2D materials \cite{Bi2021,Bi2018}. 
In details, the elemental distribution of Cu$_{\text 50}$Zr$_{\text 50}$ metallic glass is firstly visualized through continuous patterning, which is then confirmed as a Turing pattern. 
Secondly, persistent homology analysis is applied to study how temperature and interatomic potential influence the topology. Thirdly, a new criterion in addition to the negative $\Delta H_{\rm mix}$ is proposed, in order to separate the conditions for making metallic glasses from making crystals. Persistent homology analysis demonstrates that the relative magnitudes of the well depth of the interatomic potentials determines the medium-range structure of liquid alloys, which separates the products into categories including metallic glasses.

\section*{Results}
\subsection*{Modeling and its verification}
\begin{figure}[!htbp]
  \centering
  \begin{minipage}{12cm}
  \centering 
  \includegraphics[width=1\textwidth,scale=1]{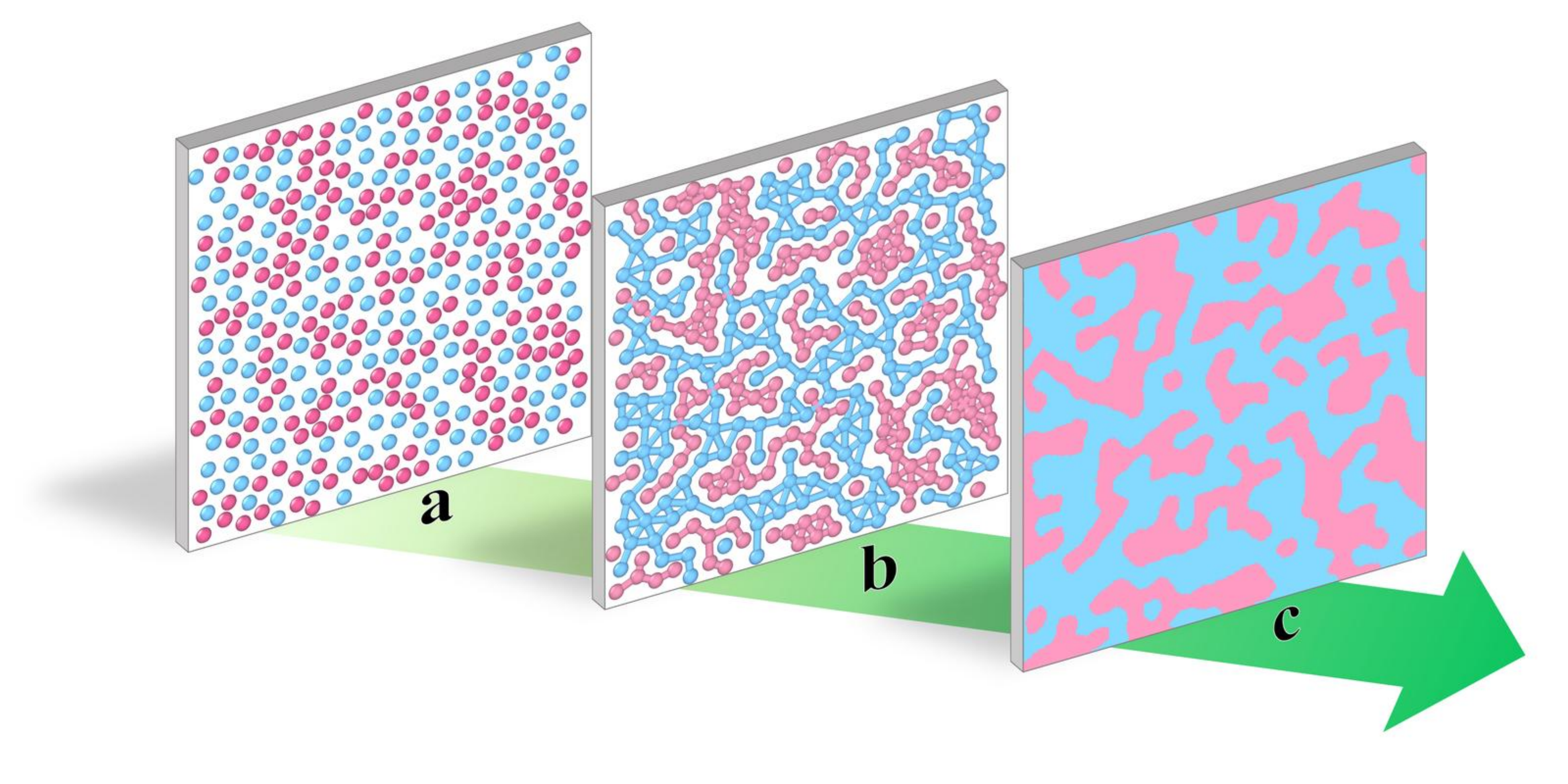} 
  \end{minipage}
  \caption{\textbf{The stucture of a 2D $\rm \bf Cu_{\text 50}Zr_{\text 50}$ liquid.} (a) Red (for Cu) and blue (for Zr) balls highlighting the atomic positions. (b) Red (for Cu-Cu) and blue (for Zr-Zr) lines highlighting atomic bonds. (c) Continuous patterning highlighting the chemical ordering. The dimensions of the representation are $5\times 5$ nm$^{\text 2}$.}
  \label{1}
  \end{figure}

A reformative microscopic depiction of metallic melts may help us grasp the CMRO. Cu$_{\text 50}$Zr$_{\text 50}$, a model MG-forming liquid, is used as an example. The model is a 45.6 $\times$ 45.6 nm$^{\text 2}$ 2D square consisting of 29600 atoms. The sample was equilibrated at 2000 K after 1 ns relaxation. A portion of the square, i.e., 5 nm $\times$ 5 nm, is cut out and shown in Fig. \ref{1}. Discrete dots emphasizing atomic positions of Cu (red dots) and Zr (blue dots) is presented in Fig. \ref{1}a. 
However, this presentation has the flaw of ignoring the atomic interactions that should take up the white space in Fig. \ref{1}a. As an alternative approach, the bondings of Cu-Cu (red lines) and Zr-Zr (blue lines) atomic pairs are demonstrated as the colored lines in Fig. \ref{1}b, from which different morphologies of Cu and Zr are revealed. Here, the Cu areas are separated, whereas the Zr areas are interconnected. 
Because Cu and Zr atoms have an equivalent atomic proportion of 50\%, such morphological differences are inherent in chemistry. Further, by filling the white space of Fig. \ref{1}a using cubic interpolation (see more details in Fig. \textcolor[rgb]{0,0,1}{S1}) , the Cu and Zr areas are highlighted in the manner of Fig. \ref{1}c emphasizing the morphological difference between Cu and Zr. In this depiction, the minutiae of atomic position, atomic size and bonding are ignored, and the gist of chemical distribution is highlighted. These morphological traits are referred to as the chemical ordering of the Cu$_{\text 50}$Zr$_{\text 50}$ liquid.

Figure \ref{1}c resembles a Turing pattern. To determine a Turing pattern, however, four requirements must be met: (i) the system consists of an activator that provides short-range positive feedback and an inhibitor that provides long-range negative feedback \cite{Bi2021,science2010,sci17,sci16}; (ii) the static image is obtained from dynamic evolution \cite{PNAS4}; (iii) different initial states have no effects on the topological properties of the final structure \cite{science2010,PNAS2009}; and (iv) the correlation length is fixed and independent on sample size \cite{np8,np9}. We now look into the validity of four criteria in the Cu$_{\text 50}$Zr$_{\text 50}$ liquid.
\begin{figure}[!htbp]
  \centering
  \begin{minipage}{12cm}
  \centering 
  \includegraphics[width=1\textwidth,scale=1]{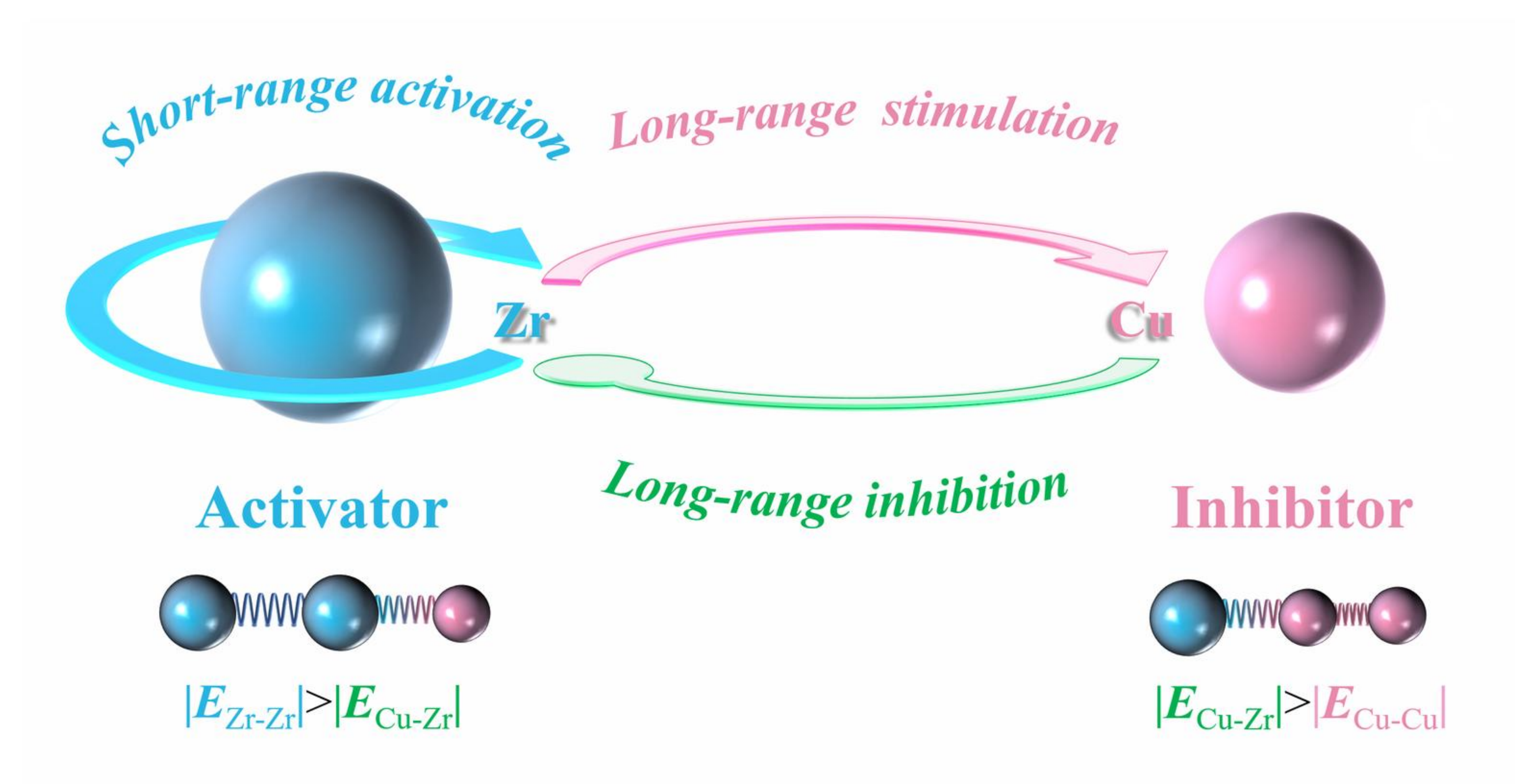} 
  \label{2C}
  \end{minipage}
  
  \caption{\textbf{The formation mechanism of Turing pattern.} Here, the activator and inhibitor are Zr (blue balls) and Cu (red balls), respectively. 
}
  \label{2}
\end{figure}

The activator and inhibitor are determined by the interatomic potentials. Depending on the valley depth of the potential functions ($\lvert E_{\text{Zr-Zr}}\rvert$ for Zr-Zr bond, $\lvert E_{\text{Cu-Zr}}\rvert$ for Cu-Zr bond and $\lvert E_{\text{Cu-Cu}}\rvert$ for Cu-Cu bond), the chemical affinity between atoms can be evaluated. For Cu-Zr, potential depth decreases in the order of Zr-Zr, Cu-Zr, and Cu-Cu. The formation of Zr-Zr and Cu-Zr bonds are active, promoting self-activation of Zr-Zr bonds and explaining the inter-connected morphology of Zr (Fig. \ref{1}c). The passive formation of Cu-Cu bonds, on the other hand, results in the long-range stimulation of Cu and the formation of Cu-rich regions. However, because the Cu-Zr bonds are energetically favored, meaning the creation of two Cu-Zr bonds at the cost of breaking Cu-Cu and Zr-Zr bonds are exothermic ($\Delta H_{\rm mix}=-23 \rm\ kJ\ mol ^{-1}$ \cite{Tak05}), the Zr region is prevented from growing bigger, while the Cu regions are separated.In other words, Cu-Zr bonds inhibit Zr areas from growing over long distances, restricting Zr-Zr activation to short distances. Here,  $\Delta H_{\rm mix}<0$ functions in the prevention of Zr-Zr region enrichment. Fig. \ref{2} summarizes the activation and inhibition schemes of Cu and Zr.

\begin{figure}[htbp]
  \centering
  \begin{minipage}{12cm}
  \centering 
  \includegraphics[width=1\textwidth,scale=0.75]{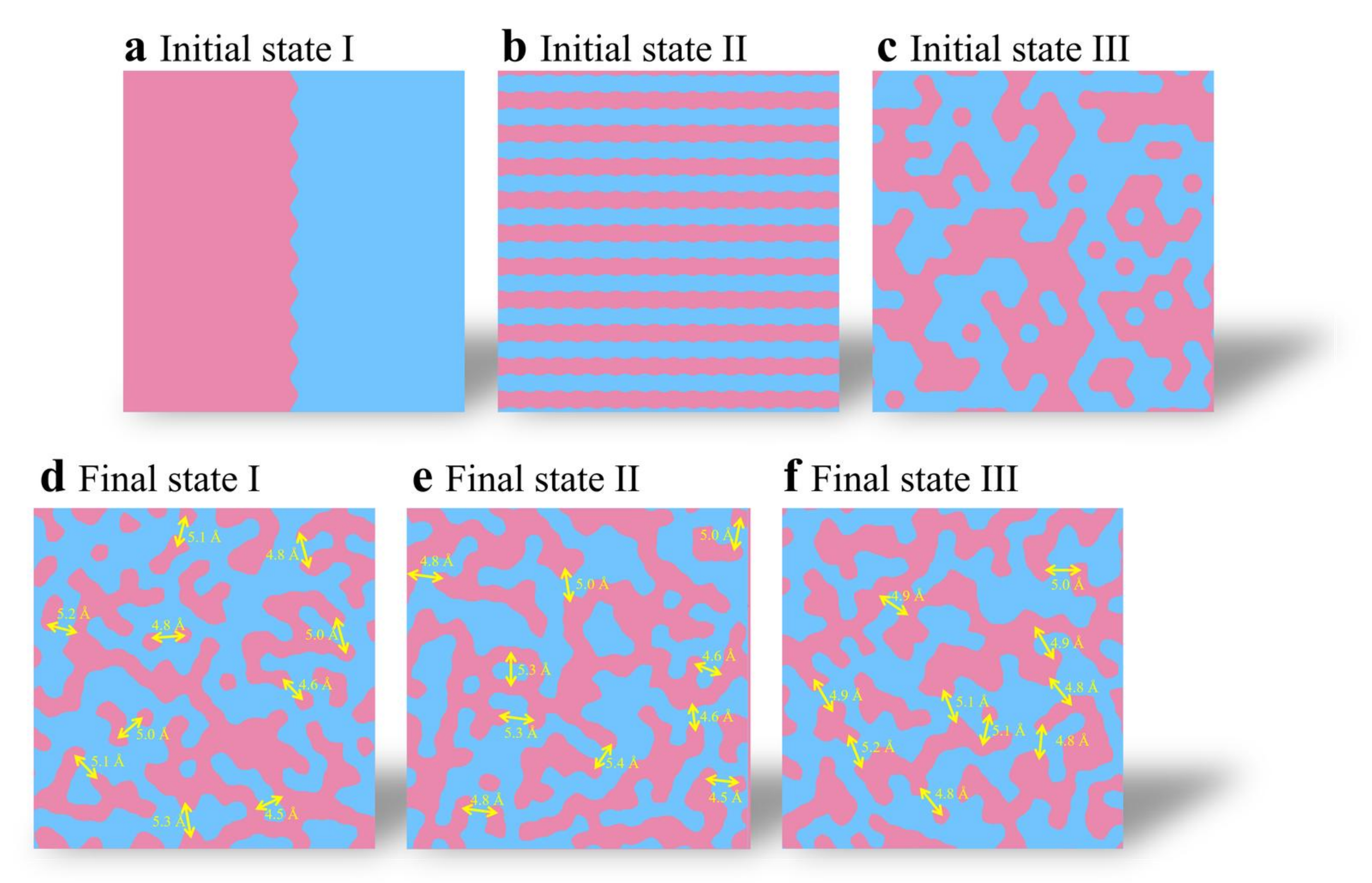} 
  \end{minipage}
  \caption{\textbf{Morphology evolution of $\rm \bf Cu_{\text 50}Zr_{\text 50}$ starting from three initial conditions including}
  (a) state I: Cu (red) and Zr (blue) are completely separated; 
  (b) state II: Cu and Zr are arranged alternately in stripes; 
  (c) state III: Cu and Zr are randomly arranged. 
  The final states are presented in (d), (e) and (f), corresponding to (a), (b) and (c) after annealing the samples at 2000 K for (d) 20 ns, (e) 1 ns and (f) 1 ns. Yellow double-ended arrows visualize the characteristic lengths of the chemical medium-range order. The dimensions of the representation are 5 $\times$ 5 nm$^{\text 2}$. }
  \label{3}
\end{figure}

A Turing pattern should be independent on the initial distribution of the activators and inhibitors. Otherwise, diffusion of the activator can create a pseudo-Turing pattern, driven by the chemical gradient of a similar pattern. The essence of a Turing pattern is the reaction between the activator and inhibitor that can erase the memory of the initial morphology. Starting from different initial distributions of Cu$_{\text 50}$Zr$_{\text 50}$, we found that the stabilized morphology is always a Turing pattern (Fig. \ref{3}). Though the final patterns are not exactly the same, the morphology of the blue and red areas are similar. Movies \textcolor[rgb]{0,0,1}{S1} and \textcolor[rgb]{0,0,1}{S2} also prove that the static Turing pattern of Cu$_{\text 50}$Zr$_{\text 50}$ is obtained by dynamic evolution. Further, the size of the model system does not matter with the pattern. In another sample with dimension of 22.7 $\times$ 22.7 nm$^{\text2}$ consisting of 7360 atoms, the resulting pattern of Movie \textcolor[rgb]{0,0,1}{S2} and Fig. \textcolor[rgb]{0,0,1}{S3} are similar to those shown in Fig. \ref{3}. Thus, the morphology of Cu$_{\text 50}$Zr$_{\text 50}$ liquid meets all the requirements of a Turing pattern.

\subsection*{Pattern analysis through persistent homology: temperature effects}

\begin{figure}[htbp]
  \centering
  \begin{minipage}{14cm}
  \centering 
  \includegraphics[width=1\textwidth,scale=0.75]{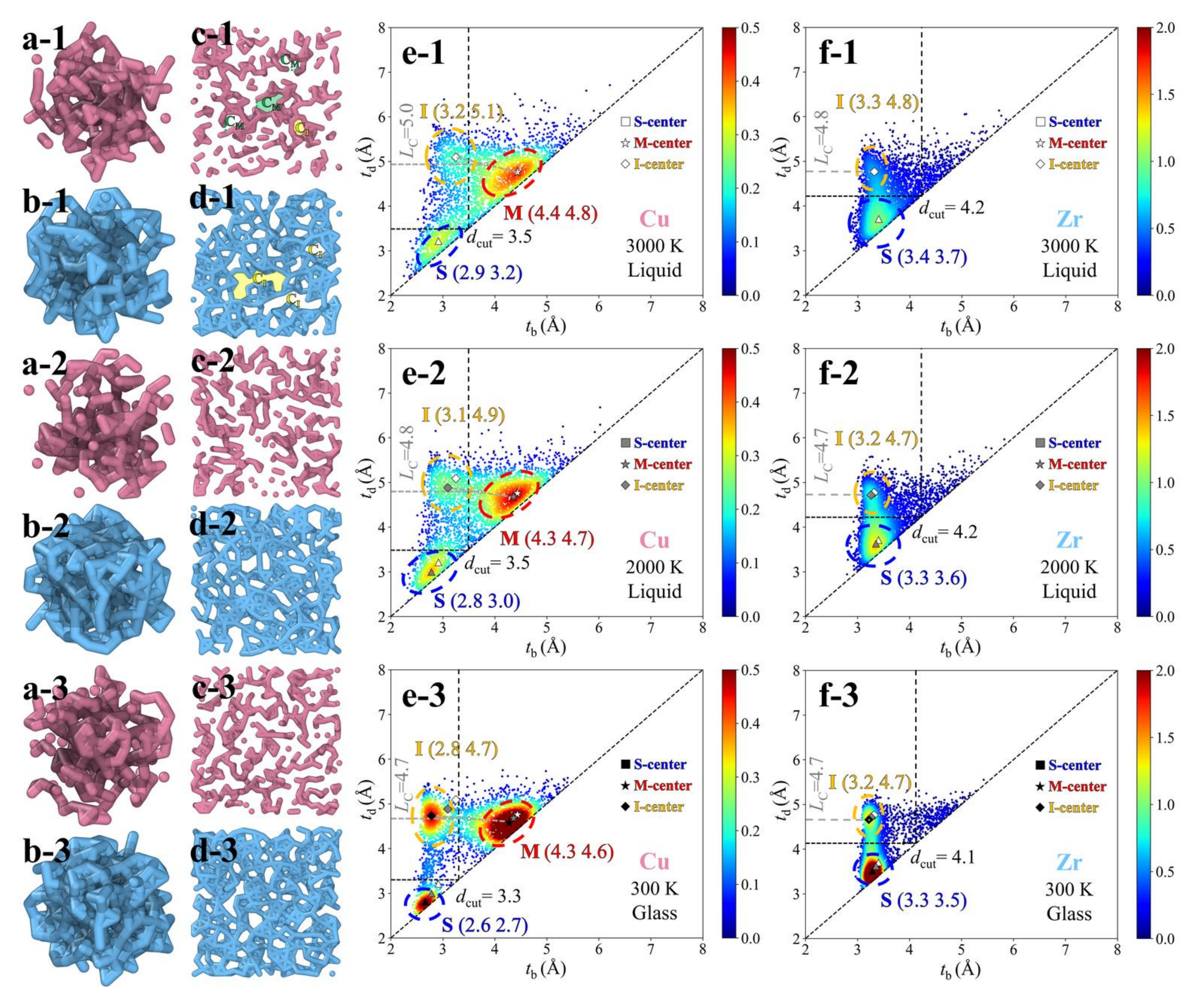} 

  \end{minipage}
  
  \caption{\textbf{Structure analysis of $\rm \bf Cu_{\text 50}Zr_{\text 50}$ in 3D at (-1) 3000 K, (-2) 2000 K and (-3) 300 K, respectively. } 3D structures of (a) Cu-Cu and (b) Zr-Zr bonds in Cu$_{\text 50}$Zr$_{\text 50}$ ($\rm 2\times 2\times 2\ nm^3$). Slice morphologies of (c) Cu-Cu and (d) Zr-Zr bonds ($\rm 5\times 5\times 0.5\ nm^3$). Persistent diagrams of (e) Cu and (f) Zr. $t_{\rm d}$, $t_{\rm b}$, $L_{\rm C}$ and $d_{\rm cut}$ denote the death time, birth time, correlation length and cut-off distance, respectivelly. }
    \label{4}
\end{figure}
The above-mentioned results of 2D Cu$_{\text 50}$Zr$_{\text 50}$ liquid provide fundamentals of a Turing pattern. We now move on to a 3D Cu$_{\text 50}$Zr$_{\text 50}$ liquid for topological analysis. 
A model system in a cube form of 17 $\times$ 17 $\times$ 17 nm$^{\text 3}$, consisting of 256000 atoms, is simulated. A cube with a side length of 2 nm is cut out, highlighting the morphologies of Cu (red) and Zr (blue) as shown in Figs. \ref{4}a and \ref{4}b. 
Here, Cu-Cu bonds are drawn when the interatomic spacing between Cu and Cu is within the cut-off distance ($d_{\rm cut}$), and $d_{\rm cut}$ is set as the locations of the first valleys in RDF (as shown in Fig. \textcolor[rgb]{0,0,1}{S2} and as collected in Table. \textcolor[rgb]{0,0,1}{S1}). The Zr-Zr bonds are drawn in the same way.
For better presentation, the bond width, i.e., the cylinder diameter, is set at 2 $\text \AA$.
A slice of 5-nm-long, 5-nm-wide and 5-$\text \AA$-thick is cut out of the 3D cube, and the resulting patterns (Figs. \ref{4}c and \ref{4}d) resemble characteristics of 2D Turing patterns (Fig. \ref{1}c). A quantitative investigation of CMRO on a Turing pattern, particularly in 3D \cite{3D}, however, necessitates persistent homology. Persistent homology is a topological method for extracting detailed properties in 3D structures \cite{2017PH}. It has been applied to extract not only structural short-range order but also structural medium-range order of oxide and metallic glasses \cite{2016PH,2020SA}. Unlike a former analysis \cite{2016PH} that neglected the element types and focused on the topology atomic structure for persistent homology analysis, the models applied here consider Cu and Zr atoms separately (Figs. \ref{4}e and \ref{4}f).

A persistent diagram (PD) is a birth-death time diagram obtained from persistent homology analysis. The birth time ($t_{\rm b}$) corresponds to the interatomic distance that forms a closed ring of bonds, while the death time ($t_{\rm d}$) is the interatomic distance that completely divides the ring by triangles (see \textbf{Methods} for more details).
Three characteristics regions are found on the PDs of Cu. The S-region is located at (2.9 $\text \AA$, 3.2 $\text \AA$) representing ($t_{\rm b}$, $t_{\rm d}$), both are smaller than $d_{\rm cut}$ of 3.5 $\text \AA$, indicating that the S-region represents the rings formed by the first-nearest neighboring Cu atoms. Here, $d_{\rm cut}$ is the abscissa of the trough of the first neighbor atom (see Fig. \textcolor[rgb]{0,0,1}{S2} and Tab. \textcolor[rgb]{0,0,1}{S1} for more details.).
The M-region shown at (4.4 $\text \AA$, 4.8 $\text \AA$) represents the rings that are formed by the second-nearest neighboring Cu atoms. Examples are provided by ``circle'' $\rm C_M$ in Fig. \ref{4}c-1. The S- and M- regions have their $t_{\rm b}$ close to $t_{\rm d}$. The I-region centered at (3.2 $\text \AA$, 5.1 $\text \AA$), reflects the rings that are formed by the first-nearest but terminated by the second-nearest neighboring Cu atoms. Examples are given by ``circle'' $\rm C_I$ in Fig. \ref{4}c-1, whose $t_{\rm d}$ is 59\% greater than its $t_{\rm b}$. Because it includes a ring with $t_{\rm d}$ $>$ $d_{\rm cut}$, both I and M regions contain medium-range information. Fig. \ref{4}c-1 provides examples of S-, M- and I-regions. The S-region reflects rings formed by the first-nearest neighboring atoms at shorter length scale; the M-region is contributed by the rings established by connecting two separated Cu regions; the I-region represents rings formed by the first-nearest neighboring atoms at longer length scale. 
For the PD of Zr (Fig. \ref{4}f-1), two distinct zones are found. Within the $d_{\rm cut}$ of 4.2 $\text \AA$, the S-region centers at (3.4 $\text \AA$, 3.7 $\text \AA$), reflecting short-range order. The I-region centers at (3.3 $\text \AA$, 4.8 $\text \AA$), reflecting medium-range order. The absence of an M-region distinguishes the PD of Zr from the PD of Cu. This indicates there are no rings produced by two separated Zr regions (Fig. \ref{4}d-1). This makes sense because all the Zr atoms are linked together. As a result, the existence of M-region established the distinction between Cu and Zr, reflecting chemical ordering at the medium range.
Since no topological regions can be found beyond 5.1 $\text \AA$ (the maximum value of coordinates of all topological regions), the CMRO as revealed by the persistent homology limits to the second-nearest neighboring atomic distance. 
The mean $t_{\rm d}$ for I and M regions was used as the $L_{\rm C}$ of CMRO. Here, the $L_{\rm C}$ that defines CMRO is similar to both Cu and Zr atoms, of 5.0 $\text \AA$ on average. The width of the stripes or the spacing between the stripes of Turing pattern of MG (Fig. \ref{1}c) visualizes the $L_{\rm C}$.

The temperature effects are further investigated. The Turing patterns exhibit similar features in both 3D (Figs. \ref{4}a and \ref{4}b) and 2D (Figs. \ref{4}c and \ref{4}d) at all three characteristic temperatures. However, the central colors and positions of the characteristic regions of their PDs differ. The central color of the regions warms up when the temperature is reduced from 3000 K to 2000 K, indicating that the liquid structure is more heterogeneous at the lower temperature. Additionally, the centers of all the regions shifts to the bottom left, indicating a compact local structure. At 300 K, the influence is increasingly pronounced as the regions highlighted by the dashed curves are nearly all filled by the central color, and the central positions are further reduced. Additionally, the boundaries separating S-, I- and M- regions are more defined in 300 K (Figs. \ref{4}e-3 and \ref{4}f-3) compared to 2000 K (Figs. \ref{4}e-2 and \ref{4}f-2). The $L_{\rm C}$ of CMRO, on the other hand, is slightly reduced to 4.7 $\text \AA$ at 300 K, in comparison with 4.8 $\text \AA$ at 2000 K and 5.0 $\text \AA$ at 3000 K. The structural ordering brought on by vitrification is reflected in a modest shortening in $L_{\rm C}$ and a narrower distribution of the characteristic local regions.

\subsection*{Pattern analysis through persistent homology: Elemental effects}
\begin{figure}[htbp]
  \centering
  \begin{minipage}{14cm}
  \centering 
  \includegraphics[width=1\textwidth,scale=0.75]{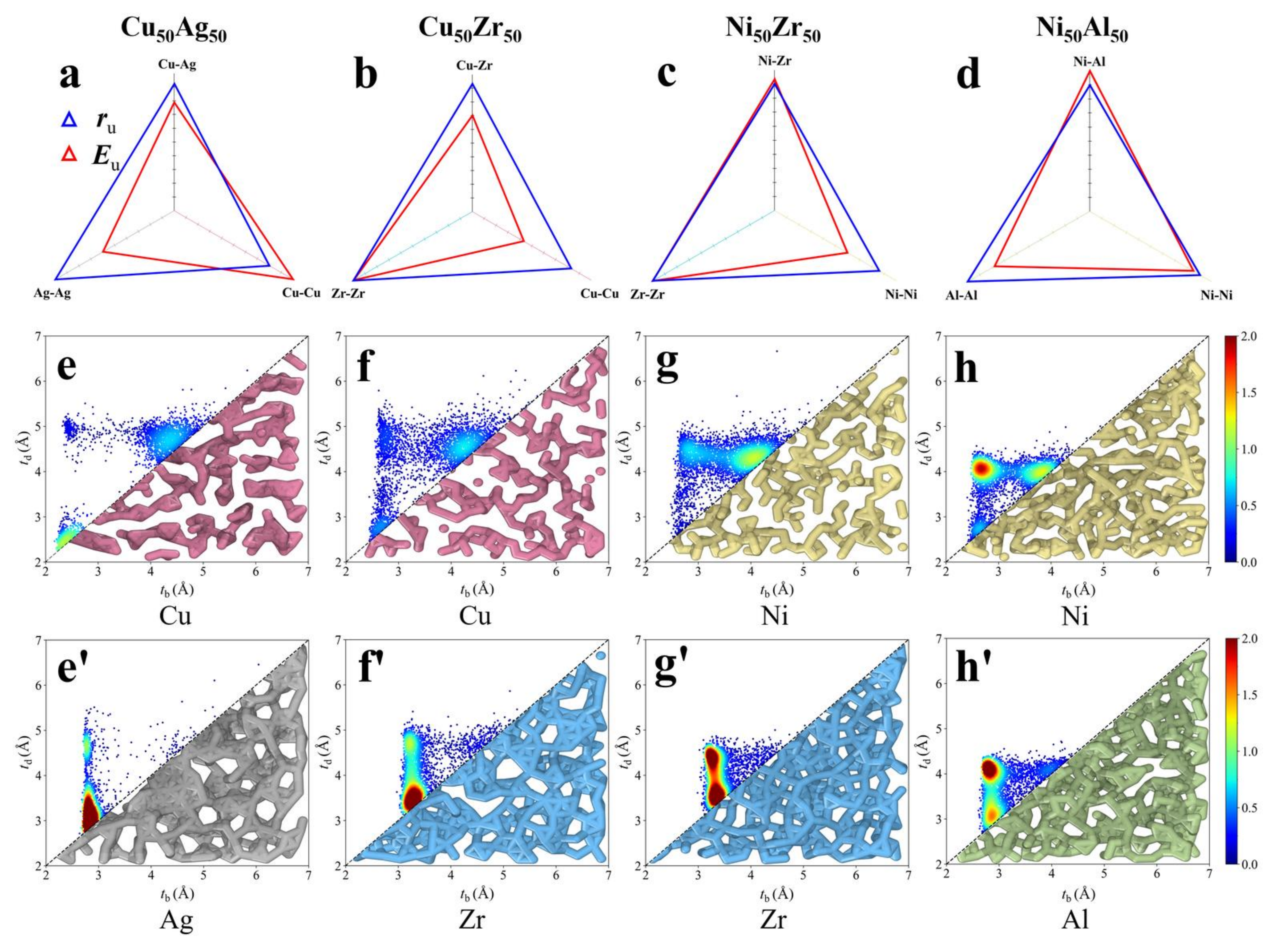} 

  \end{minipage}
  
  \caption{\textbf{Structure analysis of different binary metallic glasses in 3D at 300 K. } Radar map characterizing the potential well  of (a) Cu-Ag, (b) Cu-Zr, (c) Ni-Zr and (d) Ni-Al. $E_{\rm u}$ and $r_{\rm u}$ are normalized well energy and distance, respectively. Persistent diagrams and slice structures ($\rm 5\times 5\times 0.5\ nm^3$) of (e) Cu$_{\text 50}$Ag$_{\text 50}$, (f) Cu$_{\text 50}$Zr$_{\text 50}$, (g) Ni$_{\text 50}$Zr$_{\text 50}$ and Ni$_{\text 50}$Al$_{\text 50}$. Here, red, gray, blue, yellow and green represent Cu, Ag, Zr, Ni and Al respectively. The upper-left triangle is the persistent diagram, and the bottom-right triangle is the slice structures.}
  \label{6}
\end{figure}

The effects of elemental types are studied on four equiatomic compositions, i.e., Cu$_{\text 50}$Ag$_{\text 50}$, Cu$_{\text 50}$Zr$_{\text 50}$, Ni$_{\text 50}$Zr$_{\text 50}$ and Ni$_{\text 50}$Al$_{\text 50}$. Their radar maps (Figs. \ref{6}a, \ref{6}b, \ref{6}c and \ref{6}d) represent variations in the normalized depth of potential well ($E_{\rm u}=\frac{E}{E_{\rm min}}$) and the normalized distance of potential well ($r_{\rm u}=\frac{r}{r_{\rm max}}$). Here, $E$ and $r$  respectively are the valley energy and distance of the potential function, while $E_{\rm min}$ and $r_{\rm max}$ respectively denote the lowest values of $E$ reflecting the deepest valley and the highest $r$ among all the alloy constituents. So, a stronger atomic binding is denoted by a larger $E_{\rm u}$, and a shorter atomic bond is denoted by a smaller $r_{\rm u}$. For Cu$_{\text 50}$Ag$_{\text 50}$ (Fig. \ref{6}a) and Cu$_{\text 50}$Zr$_{\text 50}$ (Fig. \ref{6}b), the $r_{\rm u}$-triangles are comparable, but the $E_{\rm u}$-triangles are different. The Cu-Cu bond is the strongest in Cu$_{\text 50}$Ag$_{\text 50}$ and the weakest in Cu$_{\text 50}$Zr$_{\text 50}$. Due to this, the S-region in the PD of Cu in Cu$_{\text 50}$Ag$_{\text 50}$ (Fig. \ref{6}e) and the M-region in the PD of Cu in Cu$_{\text 50}$Zr$_{\text 50}$ (Fig. \ref{6}f) are both the strongest. However, the PD of Ag in Cu$_{\text 50}$Ag$_{\text 50}$ (Fig. \ref{6}e') resembles the PD of Zr in Cu$_{\text 50}$Zr$_{\text 50}$ (Fig. \ref{6}f'), both containing S- and I-regions. Despite having various $E_{\rm u}$-triangles, the similar topological properties are determined by the large potential well distance  of Ag and Zr ($r_{\rm u}$=1). Given that the S-regions of Cu and Ag are both the warmest in color among their S-, I- and M-regions, it suggests that both the activator and inhibitor of Cu$_{\text 50}$Ag$_{\text 50}$ are closely packed in topology (Figs. \ref{6}e and \ref{6}e'), forming a microscopic structure of phase separation, and creates difference from the topology of Cu$_{\text 50}$Zr$_{\text 50}$ (Figs. \ref{6}f and \ref{6}f'). This is expected by the positive $\Delta H_{\rm mix}$ of Cu-Ag\cite{Tak05} and supported by experimental evidence \cite{Chen2007,Nag2009}.
In the case of Ni$_{\text 50}$Zr$_{\text 50}$, it likewise possesses a similar $r_{\rm u}$-triangle to Cu$_{\text 50}$Zr$_{\text 50}$, but the Ni-Zr bond is stronger than Cu-Zr bond in terms of $E_{\rm u}$ (Figs. \ref{6}b and \ref{6}c). As a result, the I-region is stronger in the PDs of both Ni and Zr (Figs. \ref{6}g and \ref{6}g') than it is in the PDs of both Cu and Zr (Figs. \ref{6}f and \ref{6}f'). 
The contradiction comes from Ni-Al which has a negative $\Delta H_{\rm mix}$ \cite{Tak05}, and is a bad MG former as reported\cite{Tang2018}. The Ni-Al bond is the strongest for Ni$_{\text 50}$Al$_{\text 50}$ (Fig. \ref{6}d), therefore it tends to break Ni-Ni and Al-Al bonds. This rises the intensity of region I for both Ni and Al (Figs. \ref{6}h and \ref{6}h'), resulting in a binary crystal structure easily. Thus, the clues of glass formation resides in the intensity order of these PD regions.

\section*{Discussion}
\begin{figure}[htbp]
  \centering
  \begin{minipage}{12cm}
  \centering 
  \includegraphics[width=1\textwidth,scale=0.75]{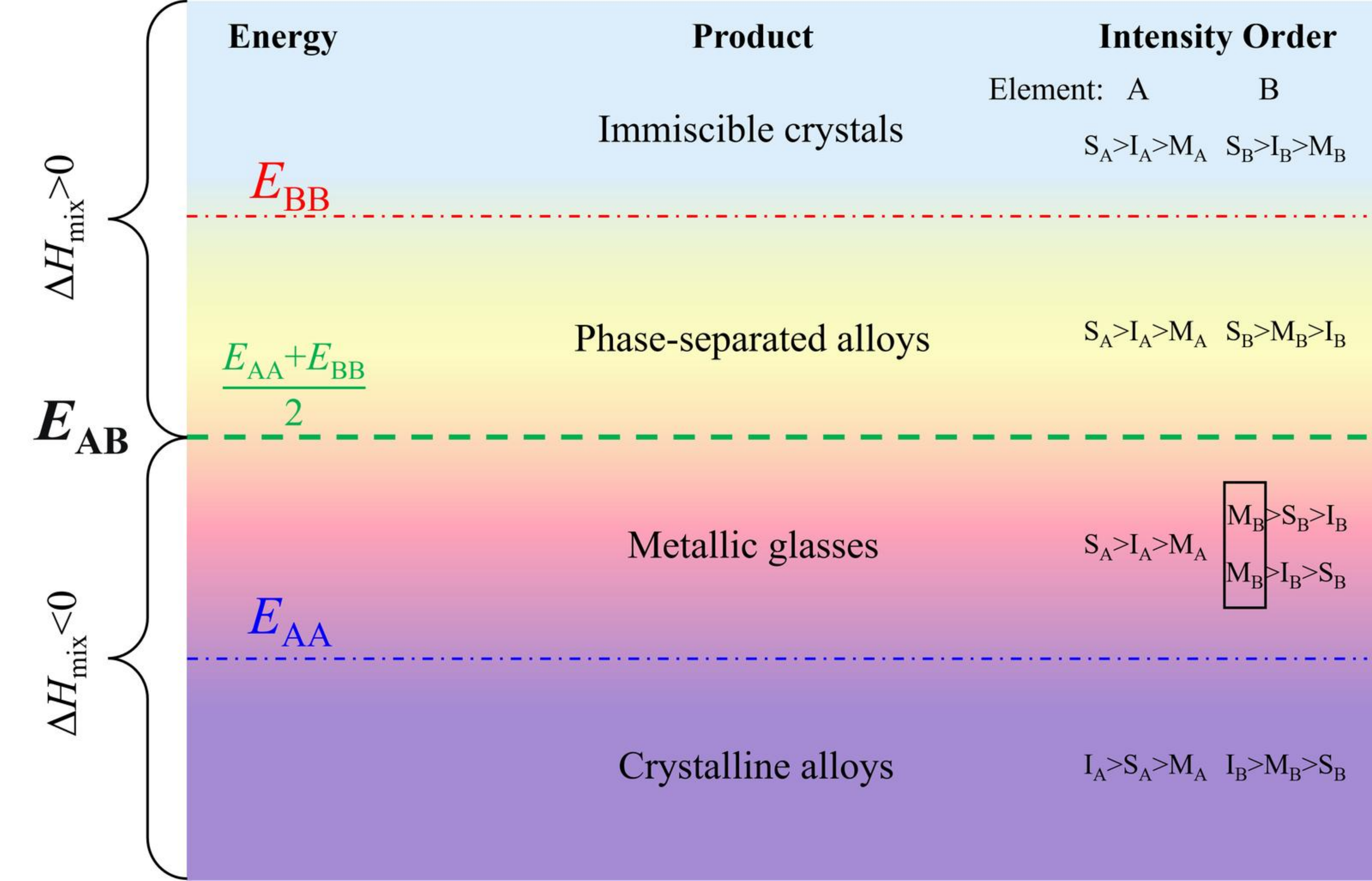} 

  \end{minipage}
  
  \caption{\textbf{Schematic diagram explaining how interatomic-potential energy depth ($E_{\rm AA}$, $E_{\rm BB}$ and $E_{\rm AB}$) and the intensity order of PD regions influence the products. } Because the heat of mixing is $\Delta H_{\rm mix}=E_{\rm AB}-\frac{(E_{\rm AA}+E_{\rm BB})}{2}$\cite{Tak2000, Hu2019}, the products are categorized into four sections,as separated by the dashed lines, according to the relative position of $E_{\rm AB}$. $\rm S_A(S_B)$, $\rm I_A(I_B)$ and $\rm M_A(M_B)$ represent the peak intensity of PD regions S, I and M of element A (or B), respectively. }
  \label{7}
\end{figure}

We now provide a schematic explanation (Fig. \ref{7}) on how chemical affinity influences the solid phase by using our data of equi-atomic binary alloys (A$_{\text 50}$B$_{\text 50}$). These alloys have similar $r_{\rm{u}}$-triangle but different combinations of interatomic potentials. Here, the chemical affinity is represented by the bond energy or the depth of the interatomic potentials ($E<0$). The absolute value of $E$ of A-A bond,  $\lvert E_{\rm AA}\rvert$, is assumed greater than that of B-B, i.e. $\lvert E_{\rm AA}\rvert>\lvert E_{\rm BB}\rvert$. 

If $\Delta H_{\rm mix}>0$ occurs when $\lvert E_{\rm AA}\rvert>\lvert E_{\rm BB}\rvert>\lvert E_{\rm AB}\rvert$, full immiscibility results in the sample. If $\Delta H_{\rm mix}>0$ is obtained at $\lvert E_{\rm AA}\rvert>\lvert E_{\rm AB}\rvert>\lvert E_{\rm BB}\rvert$, the system is partially miscible, and phase separation occurs. Here, B-B bonding is the least favored so that B-rich regions are hindered. On the other hand, A-A maintains its chemical affinity to bond with each other, which allows for A-A bond enrichment. Because A-B bond is stronger than the B-B bond, the A-rich regions are separated by the A-B bonds and surrounded by the passively formed B-rich regions, leading to phase separation. 
In the case of $\Delta H_{\rm mix}<0$, if it is acquired by $\lvert E_{\rm AA}\rvert>\lvert E_{\rm AB}\rvert>\lvert E_{\rm BB}\rvert$, MGs are formed. The A-A bonding are still preferred as A-A bond has the strongest chemical affinity, but the growth of A-rich regions are strongly inhibited due to the growing influence of A-B bonds, making crystallization from A-rich regions difficult. If, on the other hand, $\Delta H_{\rm mix}<0$ is obtained at $\lvert E_{\rm AB}\rvert>\lvert E_{\rm AA}\rvert>\lvert E_{\rm BB}\rvert$, A-B bonding is energetically preferred so that A-B bond rich regions are grown. Crystalline binary alloys are likely produced at this condition. 

The PD analysis provides a structural basis for these observations. When A and B atoms are fully immiscible, they are separatedly packed, and $\rm S_A$ and $\rm S_B$ are the strongest on their PDs. For phase separation, $\rm S_A$ and $\rm S_B$ are still the strongest because of the active enrichment of A atoms and the passive enrichment of B atoms. But, because the B-rich regions are passively formed, lacking connections to each other, a strong $M_B$ resembling linkage of isolated regions in topology is obtained, giving rise to the relationship $\rm S_B > M_B > I_B$.
In the case of MG formation, A atoms are activators whose intensity order remains unchanged with $\rm S_A$ being the strongest. However, since B atoms are bonded more strongly with A atoms, the B-rich regions are capped by A atoms, leading to a spatial distribution of discrete islands. This then makes the $\rm M_B$ the strongest in its PD. Lastly for crystalline alloys, $\rm I_A$ and $\rm I_B$ are the strongest on their respective PDs, reflecting the influence of strong A-B bonding. In this case, although $\Delta H_{\rm mix}<0$
its magnitude does not promote MG formation but instead drives crystallization. Therefore, it is crucial to maintain $\lvert E_{\rm AA}\rvert>\lvert E_{\rm AB}\rvert>\lvert E_{\rm BB}\rvert$ while keeping a negative $\Delta H_{\rm mix}$ to create MGs. This is the criteria resolved from our topological analysis. 

Finally, we highlight the broad implication this research may have. The  basis of the model is solely dependent on the binding-energy order, regardless of the kind of objects or the size of the system. The fundamental building blocks can be applied to any objects like clusters or celestial entities in addition to atoms. In theory, any systems with competitive relations might have topological patterns in common. The different ways that the constituents engage with one another are what bring order out of chaos in spatial arrangement.

\section*{Conclusion}
This paper develops a new model to visualize the chemical medium range order of metallic glasses and answers why the magnitude of negative heat of mixing is irrelevant for glass formation. By shifting the microscopic presentation from discrete to continuous patterns, the present work discloses a Turing pattern with traits of dynamic development, starting-structure irrelevance, and sample-size independence. The patterns are quantitatively analyzed by persistent homology, where the topological difference of the constituents are revealed. 
Temperature and elemental type influence the topological properties shown in the persistent diagrams. Apart from a negative heat of mixing, the bonding energy of unlike species has to be set in the middle of the bonding energy of like species so that the intensity order of the persistent-homology regions is in favor of metallic-glass formation. 




\section*{Methods}
\subsection*{MD simulations}
Molecular dynamics (MD) simulation was conducted on LAMMPS \cite{LAMMPS} software.
In this paper, the potential functions of Cu-Zr \cite{Cu-Zr}, Cu-Ag \cite{Cu-Ag}, Ni-Zr \cite{Ni-Zr} and Ni-Al \cite{Ni-Al} were used.
An isobaric and isothermal (NPT) ensemble was applied, and the system pressure was set to 0 with periodic boundary conditions. 
2D Cu$_{\text 50}$Zr$_{\text 50}$, containing 29600 atoms in a 45.6 nm $\times$ 45.6 nm square, and all 3D samples, containing 256000 atoms in cubes, were modeled. The 2D and 3D liquids were simulated by annealing the specimens at 2000 K and 3000 K for 1 ns to ensure that their liquid configurations were uniform. The 3D glass were simulated by quenching the liquids from 3000 K to 300 K at a cooling rate of $2.7\times10^{13}$ K\ s$^{-1}$.

\subsection*{Persistent Homology}
Persistent diagrams were constructed to analyze the topologies of different kind of atoms. The principles \cite{2016PH,2020SA} were to determine the birth time ($t_{\rm b}$) and death time ($t_{\rm d}$) of virtual rings. 
Due to the software's restriction on the number of atoms, 5000 atoms, carved from the center of the 3D specimen, were employed. All the persistent diagrams in this paper were obtained from the analysis of about 5000 atoms. In this work, Ripser toolkit \cite{Ripser} based on the Vietoris-Rips \cite{2019rip} was employed. In this toolkit, the virtual rings were born by connecting the virtual circles orbiting the atoms and were died when entirely divided by triangular loops. An example of $t_{\rm b}$ and $t_{\rm d}$ determination was given in Fig. \textcolor[rgb]{0,0,1}{S4}.

\section*{Acknowledgments}
This research was supported by the National Key Research and Development Plan (2018YFA0703603), the Strategic Priority Research Program of the Chinese Academy of Sciences (XDB30000000), the National Natural Science Foundation of China (51971239 and 92263103), and the Natural Science Foundation of Guangdong Province (2019B030302010). Correspondence and requests for materials should be addressed to Y.H.S. (ysun58@iphy.ac.cn). The data that support the findings of this study are available from the corresponding authors upon reasonable request.

\bibliography{prbib}

\end{document}


\begin{CJK*}{UTF8}{gbsn}
\raggedbottom
\title{Supplementary Materials for\\Turing Pattern and Chemical Medium-Range Order of Metallic Glasses}

\author{Song Ling Liu (刘松灵)\textsuperscript{1,2}}
\author{Xin Yu Luo (罗新宇)\textsuperscript{1,2}}
\author{Jing Shan Cao (曹景山)\textsuperscript{1,2}}
\author{Zhao Yuan Liu (刘召远)\textsuperscript{3}}
\author{Bei Bei Xu (许贝贝)\textsuperscript{4}}
\author{Yong Hao Sun (孙永昊)\textsuperscript{1,2,5,\Letter}}
\author{Weihua Wang (汪卫华)\textsuperscript{1,2,5}}


\affiliation{1.Institute of Physics, Chinese Academy of Sciences, Beijing, 100190, China.}
\affiliation{2.School of Physical Sciences, University of Chinese Academy of Sciences, Beijing, 100190, China.}
\affiliation{3.Shandong Computer Science Center (National Supercomputer Center in Jinan), Qilu University of Technology (Shandong Academy of Sciences), Jinan, Shandong, 250014, China.}
\affiliation{4.Shanghai Institute of Microsystem and Information Technology, Chinese Academy of Sciences, Shanghai, 200050, China.}
\affiliation{5.Songshan Lake Materials Lab, 523808, Dongguan, Guangdong, China}
\affiliation{\rm \Letter\ Corresponding authors: ysun58@iphy.ac.cn }
\date{\today}

\maketitle
\end{CJK*}
\clearpage

\section*{Supporting movies}
Movie \textcolor[rgb]{0,0,1}{S1} shows the pattern evolution of a 2D 29600-atom Cu$_{50}$Zr$_{50}$ system at 2000 K with an initial state Cu and Zr are completely separated.\\
Movie \textcolor[rgb]{0,0,1}{S2} shows the pattern evolution of a 2D 7360-atom Cu$_{50}$Zr$_{50}$ system at 2000 K, with an initial state when Cu and Zr are sliced into multilayers.

\section*{Supporting information}
A grid meshes the discrete representation of atomic positions. When the center of a Cu atom falls within a grid element, the grid element is indicated as 0, and when the center of a Zr atom falls within a grid element, it is indicated as 1. The remaining grid elements are filled by ``cubic'' interpolation of \textit{gridata} code in Matlab 2018b. After that, the numerical matrix is binarized by setting the entries to 1 or 0 depending on whether the value is equal to and larger than or smaller than 0.5. As a result, a continuous pattern is created, with red denoting elements at values of 0 and blue denoting those at values of 1.
\begin{figure}[h]
  \centering
  \begin{minipage}{12cm}
  \centering 
  \includegraphics[width=1\textwidth,scale=0.75]{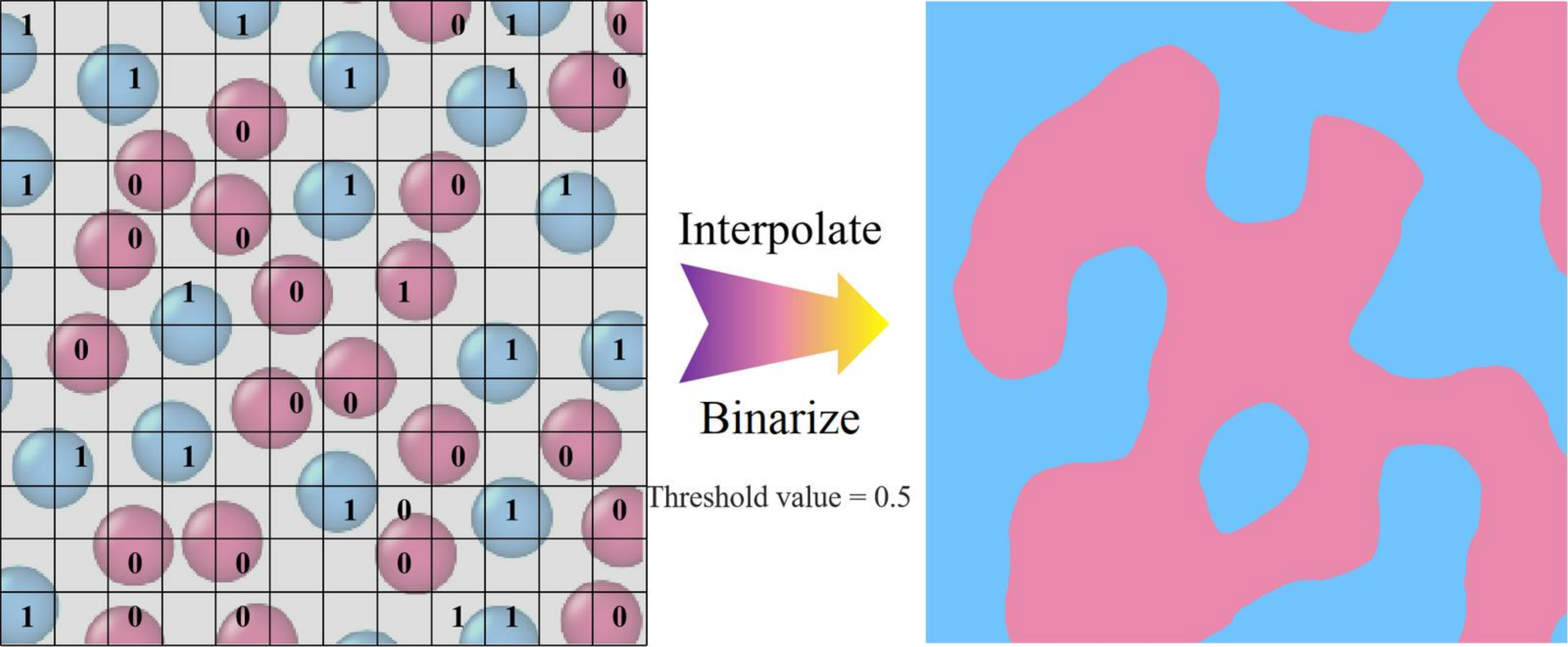} 
  \end{minipage}
\caption{\textbf{ Schematic diagrams of the continuous process of 2D $\bf Cu_{50}Zr_{50}$ structure.} }
\label{S1}
\end{figure}

\begin{figure}[htbp]
  \centering
  \begin{minipage}{12cm}
  \centering 
\includegraphics[width=1\textwidth,scale=0.75]{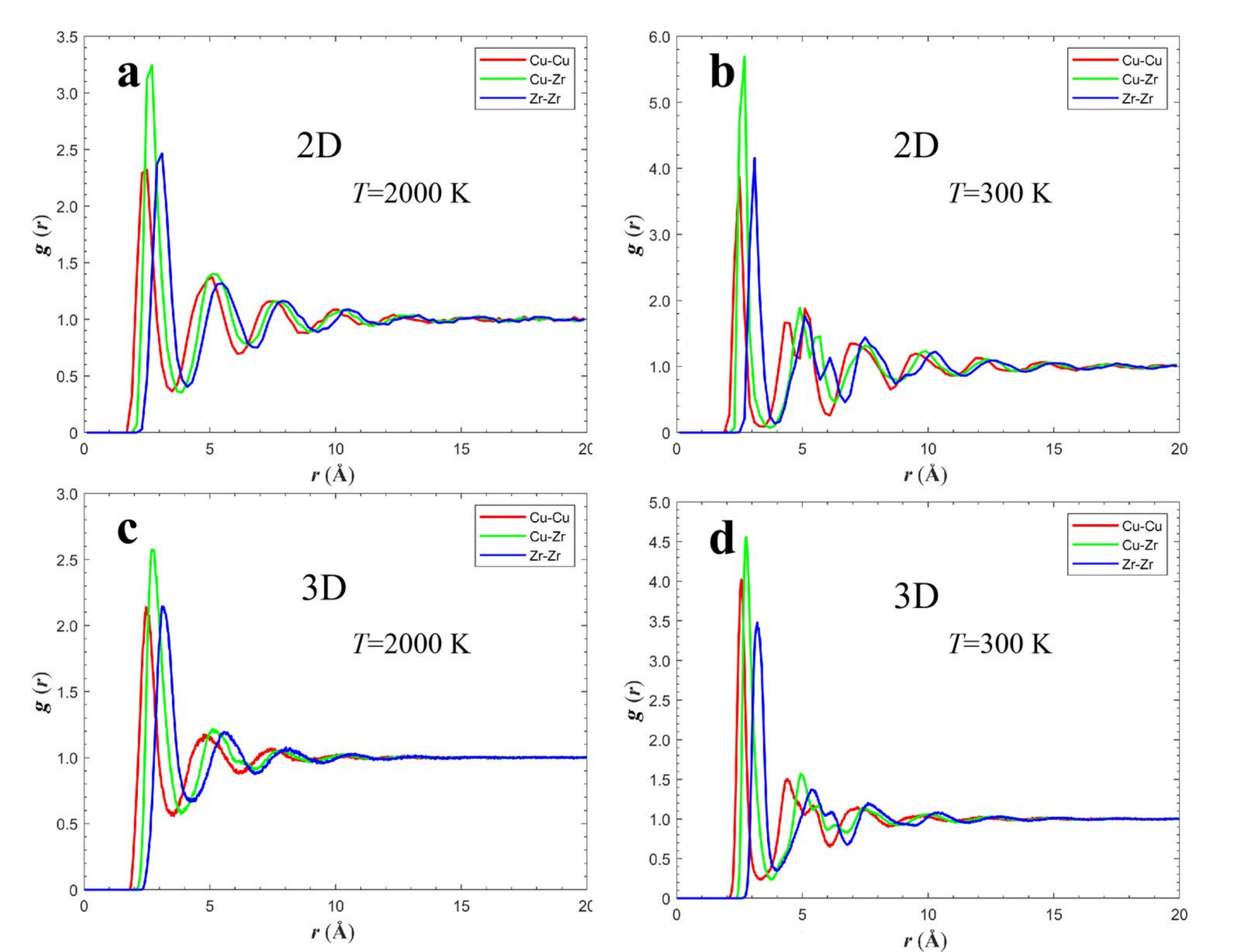} 
  \end{minipage}
\caption{\textbf{Radial distribution function (RDF) $\bm{\textsl{g}} \rm{(}\bm{\textsl{r}}\rm{)}$ of} 2D Cu$_{50}$Zr$_{50}$ at (a) 2000 K, (b) 300 K and 3D Cu$_{50}$Zr$_{50}$ at (c) 2000 K, (d) 300 K. }
\label{S2}
\end{figure}

Figure. \ref{S2} shows the $\textsl{g}(r)$ of Cu$_{50}$Zr$_{50}$ at different temperatures in 2D and 3D. 

The positions of the 1$^{\rm st}$ peak and its following valley are summarized in Table S1, where the positions of the valley defines the cut-off distance ($d_{\rm cut}$).\\ 
\begin{table}[htbp]
  \centering
  \caption*{Table S1: The locations of peaks and valleys in RDF under different circumstances.} 
  \begin{tabular}{|c|c|c|c|c|c|c|}
  \hline
  State & \multicolumn{2}{c|}{\textcolor[rgb]{ 1,  0,  0}{\textbf{Cu-Cu}} ($\rm \AA$)} & \multicolumn{2}{c|}{\textcolor[rgb]{ 0,  0,  1}{\textbf{Zr-Zr}} ($\rm \AA$)} & \multicolumn{2}{c|}{\textcolor[rgb]{ 0,  1,  0}{\textbf{Cu-Zr}} ($\rm \AA$)} \bigstrut\\
  \hline
  \textbf{2D} & peak  & valley & peak  & valley & peak  & valley \bigstrut\\
  \hline
  300 K & 2.5   & 3.4   & 3.0   & 4.0   & 2.6   & 3.8  \bigstrut\\
  \hline
  2000 K & 2.4   & 3.5   & 3.0   & 4.1   & 2.6   & 3.8  \bigstrut\\
  \hline
  \textbf{3D} & peak  & valley & peak  & valley & peak  & valley \bigstrut\\
  \hline
  300 K & 2.6   & 3.3   & 3.2   & 4.0   & 2.8   & 3.8  \bigstrut\\
  \hline
  2000 K & 2.5   & 3.5   & 3.2   & 4.2   & 2.7   & 3.8  \bigstrut\\
  \hline
  3000 K & 2.5   & 3.5   & 3.2   & 4.2   & 2.7   & 3.8  \bigstrut\\
  \hline
  \end{tabular}%
  \label{tS1}%
\end{table}%

\begin{figure}[htbp]
  \centering

  \begin{minipage}{8cm}
  \centering 
  \includegraphics[width=1\textwidth,scale=0.75]{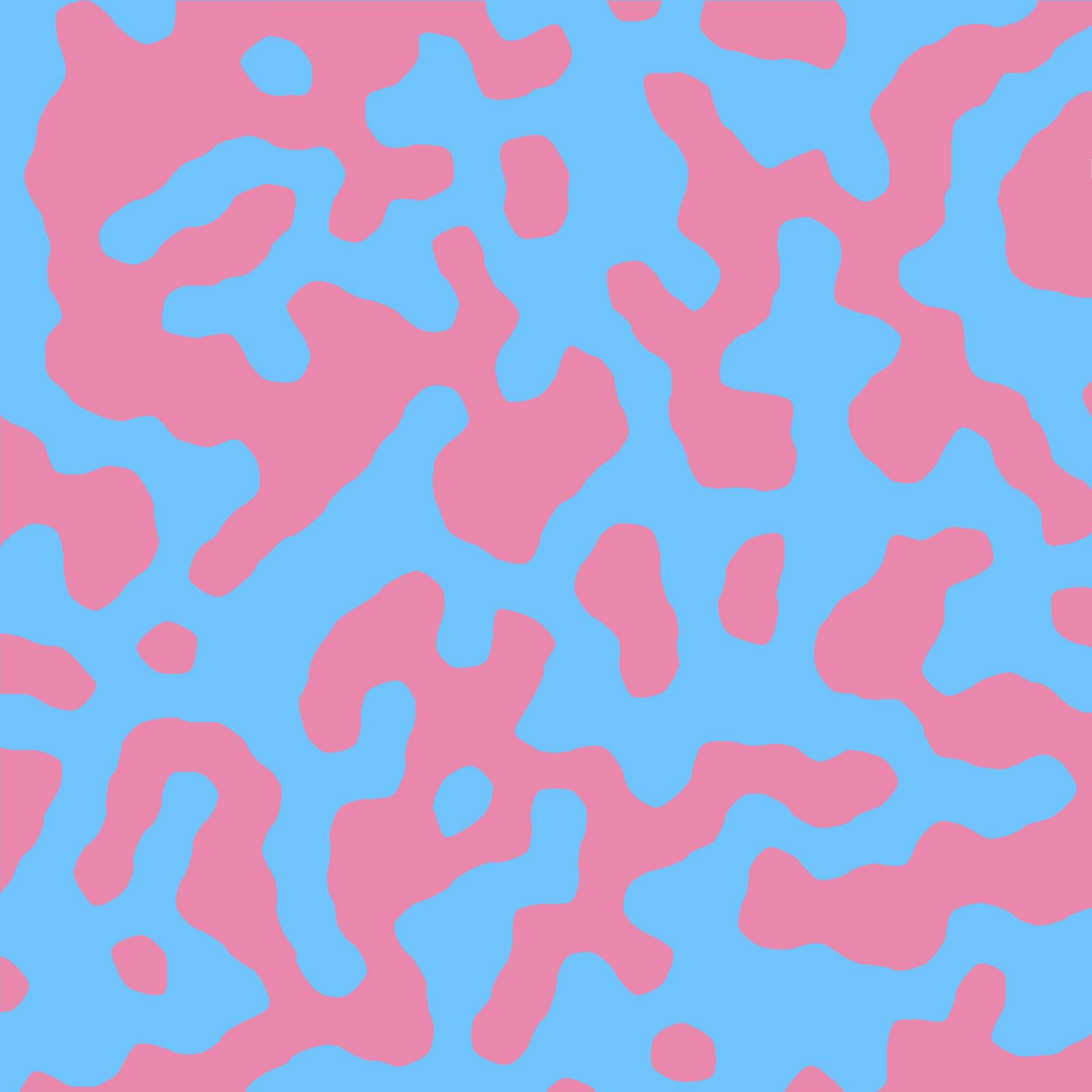} 
  \end{minipage}

\caption{Turing structure in a 7360-atom Cu$_{50}$Zr$_{50}$ system after annealing it at 2000 K for 1 ns, which is consistent with that of 29600-atom system after annealing it at 2000 K (Fig. 3), indicating that the size of the system does not affect the morphology.}
\label{S4}
\end{figure}



\begin{figure}[h]
  \centering
  \begin{minipage}{10cm}
  \centering %
  \includegraphics[width=1\textwidth,scale=0.75]{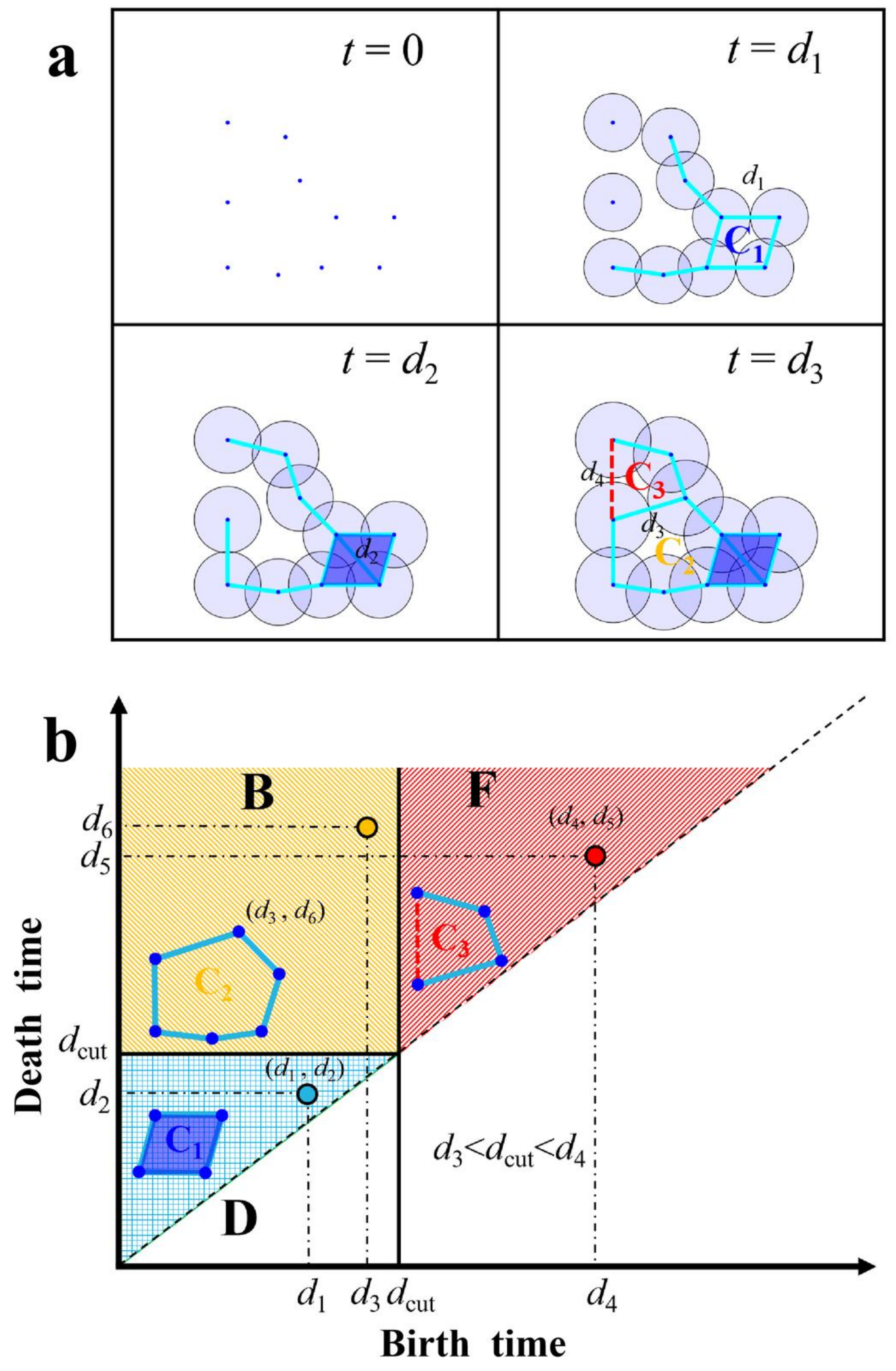} %
  \end{minipage}
  
\caption{\textbf{Schematic diagram of persistent homology.} (a) The schematic explaining the birth and death of the rings. Here, $d$ is the interatomic distance. (b) Persistent diagram of (a). The blue, yellow and red area represent the death, birth and future regions, respectively. }
\label{S5}
\end{figure}

Time ($t$) is visualized by the diameters of the circle centering each atom. When two circles are touched, a virtual bond is establish with the bond-length equal to the diameters. In this work, Ripser toolkit\cite{Ripser} based on the Vietoris-Rips\cite{2019rip} was used. When the virtual bond is connected into a polygon (quadrilateral and above), a ring is born, this time is recorded as birth time ($t_{\rm b}$). When the polygon is completely split into a combination full of triangles, the ring is dead, and this time is denoted as death time ($t_{\rm d}$). For example, as shown in Fig. \ref{S5}a, the $\bf C_1$ ring consisting of four neighboring bonds is created at $t = d_1$, yet ended at $t = d_2$ because its ring structure is completely split into triangular loops. 

Figure \ref{S5}b is called  persistent diagram (PD), which takes $t_{\rm b}$ and $t_{\rm d}$ obtained in Fig. \ref{S5}a as abscissa and ordinate, respectively. In Fig. \ref{S5}b, a time $t=d_{\rm cut}$ is selected, which divides the PDs into 3 dashed regions: blue, yellow and red, so called region D (death), region B (birth) and region F (future) respectively.
When $t=d_{\rm cut}$, as shown in Fig. \ref{S5}a, the dead ring appears in region D, the undead ring appears in region B, and the ring about to be born appears in region F. 
For example, when $d_3\leq d_{\rm cut}<d_4$, the connection status of the virtual bond is the same as that of $t=d_3$. 
At this time, in Fig. \ref{S5}a, $\bf C_1$ has been dead, $\bf C_2$ has been born but undead, and $\bf C_3$ has not yet been born. $t_{\rm b}$ and $t_{\rm d}$ of $\bf C_1$, $\bf C_2$ and $\bf C_3$ are drawn as blue, yellow and red points, respectively, in Fig. \ref{S5}b, which fill into region D, B and F, respectively.

\clearpage

\bibliography{prbib}